\documentclass[a4paper,12pt]{iopart}
\usepackage{graphicx}
\usepackage{amssymb}
\def\bea {\begin{eqnarray}}
\def\eea {\end{eqnarray}}
\def\be {\begin{equation}}
\def\ee {\end{equation}}
\def\nn {\nonumber}

\begin{document}

\title{The effect of neutron skin on inclusive prompt photon production in Pb~+~Pb collisions at the LHC}
\title[The effect of neutron skin....]{}

\author{Somnath De\footnote[1]{somvecc@gmail.com}}

\address{Institute of Physics, Bhubaneswar 751005, India}

\date{\today}

\begin{abstract}
Recent experiments on lead (\textrm{$Pb_{82}^{208}$}) nuclei have observed the celebrated phenomenon of neutron skin-thickness 
of low energy nuclear physics. The skin-thickness provides a measure of extension of spatial distribution of neutrons inside the atomic 
nucleus than protons. 
We have studied the effect of 
neutron skin-thickness on inclusive prompt photon production in Pb~+~Pb collisions at the Large Hadron Collider energies. 
We have calculated the \textquoteleft central-to-peripheral ratio\textquoteright ($R_\textrm{cp}$) of prompt photon 
production with and without accounting for neutron skin effect. The neutron skin causes a characteristic enhancement in the ratio, 
in particular at forward rapidity, which is distinguishable in our calculation. However a very precise direct photon measurement up to 
large transverse momenta would be necessary to constrain the feature in experiment. 
\end{abstract}

\noindent{\it Keywords}: Neutron skin, Prompt photon, Heavy-ion collisions.
\pacs{21.10.Gv, 12.38.-t, 25.75.Cj}
\maketitle
\section{Introduction}
Measurement of proton (charge) and neutron (matter) distributions inside atomic nuclei is the one most explored problem in low energy nuclear physics. 
The proton distributions are known from the electron scattering experiments~\cite{charge-rad} with high accuracy whereas the neutron distributions 
are known from the hadron scattering experiments~\cite{mass-rad} with less precision. Such experiments on neutron-rich nuclei reveal that the 
root mean square (rms) radius of neutron distributions is larger than the rms radius of charge (proton) distributions. The difference often 
referred as \textquoteleft neutron skin-thickness \textquoteright. Two recent experiments on lead (\textrm{$Pb_{82}^{208}$}) nuclei have devised 
different methods namely, parity violating electron scattering~\cite{prl12} and coherent photo-production of pions~\cite{prl14} to measure 
the parameter. Both of them have reported nearly similar value of neutron skin-thickness parameter with very good accuracy. Thus the 
neutrons are found to be more populated towards the surface than protons, for Pb nuclei. In recent years, several experiments are to be designed 
for other neutron rich spherical nuclei for which the parameter had been predicted~\cite{agrawal}. The parameter is closely related to nuclear 
symmetry energy~\cite{nskin-Warda} thus holds unique importance 
for astrophysical phenomena~\cite{nskin-astro}, equation of state (EoS) of neutron matter~\cite{nskin-eos}, etc. The observation has further 
implication in the context of Pb ion collisions at the Large Hadron Collider (LHC) energies. 
The number of neutrons can be found larger than the number of protons inside the overlap zone, in case of ultra-peripheral collisions (i.e., impact 
parameter of collision is comparable to diameter of colliding nuclei). Thus the phenomena may lead to intriguing effect in direct 
(prompt) photon production in such collisions.

The direct photon has hold the legacy of most cleanest probe of strongly interacting matter created in relativistic heavy ion collisions 
since the days of Super Proton Synchrotron (SPS) experiment at CERN. The term \textquotedblleft direct\textquotedblright refers to the class 
of photons which do not come from 
the decay of hadrons (e.g., $\pi^0$, $\eta$). Several sources of direct photons have been proposed in theory which have constituted
the entire transverse momentum ($p_T$) spectrum recorded in experiments~\cite{Paul,dks}. The prompt direct photons are sensitive to the parton 
distributions of the beam nucleon~\cite{Ichou,Sde} and isospin of the projectile and target nucleon~\cite{dks}. Their contribution mostly dominate 
the high $p_T$ region of the spectrum. 
The thermal direct photons contain information about the temperature, thermalisation time of the dense partonic medium formed in 
these collisions~\cite{Paul,dks2}. 
They have chief contribution at low $p_T$ part of the spectrum. 
Another class of photons known as \textquotedblleft jet-thermal\textquotedblright photons, have shown important contribution at 
the intermediate $p_T$ region of the spectrum. These photons are found to be sensitive to the temperature of the quark-gluon medium 
and energy loss of quarks in the medium~\cite{Fries,Sde2}.
The prompt direct photons are often categorized in two types; inclusive and isolated.
In this article we are mainly focused to inclusive prompt photons, emitted in hard 
scatterings during the initial phase of collision. We shall refer to inclusive prompt photons as prompt photons henceforth.
\section{Formalism}
The production cross-section of prompt photons in elementary hadron-hadron collisions in the factorisation regime of 
quantum chromodynamics (QCD) is described as~\cite{Arleo}:
\bea
\frac{d^2\sigma^{\gamma}}{d^2p_T dy} 
= \sum_{i,j} \int \,dx_1  f^i_A(x_1,Q_f^2) \! \int \,dx_2  f^j_B(x_2,Q_f^2) \nn\\
\times \sum_{c=q,g} \int \, \frac{dz}{z^2}  \,
\frac{{d\sigma}_{ij \rightarrow cX}(x_1,x_2;Q^2_R)}{d^2p^{c}_T dy_c} D_{c/\gamma}(z,Q^2_F),
\label{cross}
\eea
where $f^i_A(x_1,Q_F^2)$ is the parton distribution function (pdf) of the $i$-th parton,
carrying momentum fraction $x_1$ of the hadron A and similarly $f^j_B(x_2,Q_F^2)$ for hadron B. $D_{c/\gamma}(z,Q_F^2)$ 
is the parton-to photon vacuum fragmentation probability defined
at the momentum fraction $z=p_\gamma/p_c$. When the photon is emitted \textquotedblleft directly\textquotedblright in a 
hard scattering, the fragmentation function reduces to $\delta(1-z)$.
The parton-parton cross-section $\sigma^{ij\to cX}(x_1,x_2,Q_R^2)$, includes all the leading order 
$\mathcal{O}(\alpha_s^2)$ (for c$= \gamma$) and next-to-leading order $\mathcal{O}(\alpha_s^3)$ processes (for c$\neq \gamma$)
in strong coupling ($\alpha_s$). $p_T$ is the transverse momentum and $y$ is the rapidity of the photon.

In case of nucleus-nucleus (AA) collisions, we replace the free nucleon pdf by isospin averaged nuclear pdf in the above formula:
\be
f^i_A(x,Q^2)= R_A(x,Q^2)\{\frac{Z}{A}f^i_p(x,Q^2)+ \frac{N}{A}f^i_n(x,Q^2)\},
\label{npdf}
\ee
where $f^i_p$, $f^i_n$ are the free proton and neutron pdf, connected by isospin symmetry. $R_A(x,Q^2)$ denotes the nuclear modification 
to the pdf, which arises due to many body effects in high energy QCD~\cite{shadow}. Z is the proton number, N is the neutron number and 
A ($= Z~+~N$) is the mass number 
of the target and projectile nuclei. We have used EKS98 parameterization~\cite{eks98} of nuclear shadowing function in this work.
 
Now the effective neutron and proton numbers are varying from central to peripheral AA collisions because of the change in number of 
participant nucleons ($N_{part}(b)$) in the overlap zone. However the proton-to-neutron number ratio (or vice versa) remains same as 
of the original nuclei. Thus we write the effective proton and neutron number associated with collisions of impact parameter b:
\be
Z_{eff}= \frac{Z}{A}\frac{N_{part}(b)}{2}, \hspace{0.4cm} N_{eff}= \frac{N}{A}\frac{N_{part}(b)}{2}.
\label{zeff}
\ee
The method has been advocated earlier in~\cite{Rupadi} to calculate centrality dependent direct photon production at CERN SPS energies. The number of 
participant nucleons for collisions of impact parameter b are calculated using optical Glauber model~\cite{Miller}:
\bea
N_{part}(b)&=& \int \,d^2s T_A(s_1)\{1-(1-\frac{\sigma T_B(s_2)}{B})^B\} \nn\\
&+& \int \,d^2s T_B(s_2)\{1-(1-\frac{\sigma T_A(s_1)}{A})^A\},
\label{Npart-old}
\eea
where $s_{1,2}= (x\pm b/2, y)$. $T_A$, $T_B$ are the nuclear thickness functions, provide the density of nucleons in the transverse (x,y) plane.

The Woods-Saxon density distribution of nucleons in the nucleus of mass number A writes as:
\be
\rho(r)= \frac{\rho_0}{1+exp(\frac{r-R}{c})},
\label{WS1}
\ee
\begin{figure}
\begin{center}
\includegraphics[width= 7.7 cm,clip= true]{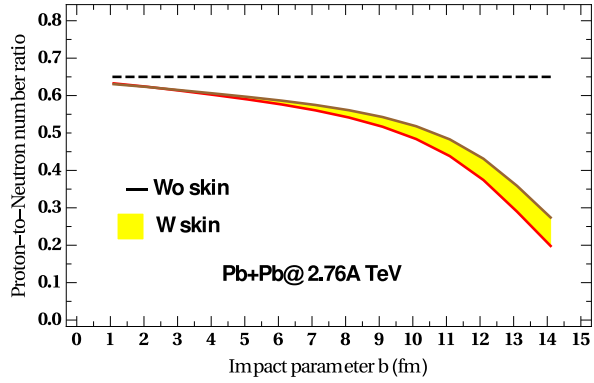}
\end{center}
\caption{(Color online) The variation of effective proton-to-neutron number ratio with impact parameter (b) of Pb~+~Pb collisions at 
$\sqrt{s_{NN}}= 2.76$ TeV. The dashed line corresponds to \textquoteleft Without \textquoteright and the 
solid lines are correspond to \textquoteleft With \textquoteright neutron skin effect. The band depicts the 
uncertainty in the ratio, due to error in the parameters of neutron density distribution~\cite{prl14}.}
\label{ratio}
\end{figure}
where $R$ is the half-density radius and $c$ is the diffuseness parameter of nuclear surface. $\rho_0$ is the nuclear saturation density 
which is fixed by the normalization criterion;
\be
\int \,d^3r \rho(r) = A,
\ee
and the nuclear thickness function:
\be
T_A(x,y) = \int \,dz \rho(r).
\ee
In the above description, the protons and neutrons are assumed to follow same distribution(Eq.~\ref{WS1}) inside the nucleus.
However the experiments on Pb nuclei~\cite{prl12, prl14} suggests that proton and neutron distributions are described by two 
different Woods-Saxon distributions:
\be
\rho_A^{p,n}(r)= \frac{\rho_0^{p,n}}{1+exp(\frac{r-R_{p,n}}{c_{p,n}})}.
\label{WS2}
\ee
The parameters are given by; $R_p= 6.680$ fm, $c_p= 0.447$ fm, $R_n= 6.70\pm 0.03$fm, $c_n= 0.55\pm 0.01$fm respectively~\cite{prl14}. The saturation 
densities of the distributions are obtained from the criteria:
\be
\int \,d^3r \rho_A^p (r) = Z = 82, \hspace{0.4cm} \int \,d^3r \rho_A^n (r) = N = 126.
\ee
Thus we can write,
\bea
\int \,dz \rho(r) &=& \int \,dz \rho_A^p(r)~+~\int \,dz \rho_A^n(r), \nn\\
T_A (s_1) &=& T_A^p (s_1)~+~T_A^n(s_1), \nn\\
T_B (s_2) &=& T_B^p (s_2)~+~T_B^n(s_2).
\eea
Inserting the above expressions of nuclear thickness functions in Eq.~\ref{Npart-old}, we get
\bea
N^{AB}_{part}(b)&=& \int \,d^2s [T_A^p(s_1)~+~T_A^n(s_1)]\{1-(1-\frac{\sigma [T_B^p(s_2)+T_B^n(s_2)]}{B})^B\} \nn\\
&+& \int \,d^2s [T_B^p(s_2)~+~T_B^n(s_2)]\{1-(1-\frac{\sigma [T_A^p(s_1)~+~T_A^n(s_2)]}{A})^A\}
\label{Npart-new}
\eea
The Eq.~\ref{Npart-new} is expanded in the power of A (and B) and the arising terms are grouped into four different categories. The each 
component is expressed in general as;
\bea
N^{pp}_{part}(b) &=& \int \,d^2s T_A^p(s_1)\{1-(1-\frac{\sigma T_B^p(s_2)}{Z_B})^{Z_B}\} \nn\\
&+& \int \,d^2s T_B^p(s_2)\{1-(1-\frac{\sigma T_A^p(s_1)}{Z_A})^{Z_A}\},
\eea
\bea
N^{nn}_{part}(b) &=& \int \,d^2s T_A^n(s_1)\{1-(1-\frac{\sigma T_B^n(s_2)}{N_B})^{N_B}\} \nn\\
&+& \int \,d^2s T_B^n(s_2)\{1-(1-\frac{\sigma T_A^n(s_1)}{N_A})^{N_A}\},
\eea
\bea
N^{pn}_{part}(b) &=& \int \,d^2s T_A^p(s_1)\{1-(1-\frac{\sigma T_B^n(s_2)}{N_B})^{N_B}\} \nn\\
&+& \int \,d^2s T_B^n(s_2)\{1-(1-\frac{\sigma T_A^p(s_1)}{Z_A})^{Z_A}\},
\eea
\bea
N^{np}_{part}(b) &=& \int \,d^2s T_A^n(s_1)\{1-(1-\frac{\sigma T_B^p(s_2)}{Z_B})^{Z_B}\} \nn\\
&+& \int \,d^2s T_B^p(s_2)\{1-(1-\frac{\sigma T_A^n(s_1)}{N_A})^{N_A}\}.
\eea
Thus we can write approximately,
\be
N^{AB}_{part}(b) \approx N^{pp}_{part}(b)~+~N^{nn}_{part}(b)~+~N^{pn}_{part}(b)~+~N^{np}_{part}(b).
\label{Npart-new2}
\ee
The Eq.~\ref{Npart-new2} is describing total number of participant nucleons on introducing the effect of neutron skin-thickness.
The contributions from proton-proton, neutron-neutron, neutron-proton (or vice versa) inelastic collisions can be factorized now.
We shall refer the first two terms as \textquoteleft pure\textquoteright and the rest two terms as \textquoteleft mix\textquoteright contribution.
In case of symmetric nuclear collisions it can be noted that $N^{pn}_{part}(b) = N^{np}_{part}(b)$.\\
We have adopted the following ansatz for symmetric collisions:
\bea
Z_{pure} &=& \frac{N^{pp}_{part}(b)}{2}, \hspace{0.4cm} N_{pure}= \frac{N^{nn}_{part}(b)}{2}; \\
Z_{mix} &=& \frac{Z_{pure}}{A_{pure}} N^{pn}_{part}(b), \hspace{0.4cm} N_{mix}= \frac{N_{pure}}{A_{pure}} N^{pn}_{part}(b),
\eea
\begin{figure}
\begin{center}
\includegraphics[width= 8.0 cm,clip= true]{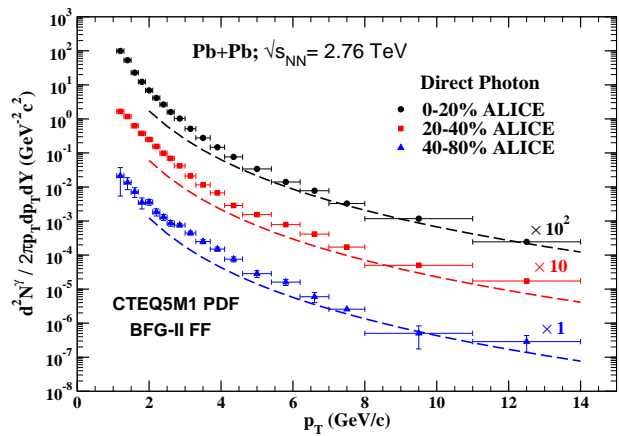}
\end{center}
\caption{(Color online) The prompt photon invariant yield is plotted as a function of $p_T$ for Pb~+~Pb collisions at 
$\sqrt{s_{NN}}= 2.76$ TeV for three collision centralities namely, 0-20\%, 20-40\%, 40-80\%. 
The results are compared with the direct photon measurement by ALICE Collaboration~\cite{Alice}.}
\label{dirphot}
\end{figure}
where $A_{pure}=Z_{pure}+N_{pure}$. The above equations are describing the proton and neutron number associated with each kind of 
inelastic collisions between the nucleons in an event. However the effective proton-to-neutron number ratio will depend on 
\textquoteleft pure\textquoteright terms only.  
We shall use these numbers in Eq.~\ref{npdf} while calculating prompt photon production with neutron skin-thickness. 
The variation of effective proton-to-neutron number ratio with impact parameter b for Pb~+~Pb collisions at 2.76A TeV is 
depicted in Fig.~\ref{ratio}.  
In case of most central collisions (i.e. small b), the ratio is closer to $Z/N(\sim0.65$) value of the nucleus. 
It is seen to deviate gradually from the 
constant value for peripheral collisions (i.e. large b) as the number of neutrons increases than the number of protons in the overlap zone. 
The uncertainty (shown by the band) in the ratio arises due 
to error in the measurement of $R_n$ and $c_n$. The result has been found qualitatively similar to an earlier work which had calculated 
the effective proton-to-neutron ratio via different method~\cite{pakkunen}.
\section{Results}
In the earlier section we found that the effect of neutron skin-thickness becomes prominent for most peripheral collisions. Thus 
\textquoteleft central-to-peripheral ratio\textquoteright ($R_\textrm{cp}$) of particle production would be the best suited observable 
to study the consequences of neutron skin on electromagnetic probes. We have calculated $R_\textrm{cp}$ of prompt photon production 
for Pb~+~Pb collisions at the LHC energies namely, 2.76 TeV and 5.5 TeV 
per nucleon pair. It is defined as the ratio of prompt photon invariant cross-section in central and peripheral AA collisions, scaled to 
per nucleon-nucleon collision:
\be
R_\textrm{cp} (p_T, b)=\frac{d^2\sigma^{\gamma}/ d^2p_T dy_{\hspace{0.1cm}\textrm{central}}}
{d^2\sigma^{\gamma}/ d^2p_T dy_{\hspace{0.1cm}\textrm{peripheral}}}.
\label{Rcp}
\ee
\begin{figure}[t]
\begin{center}
\includegraphics[width= 7.7cm,clip= true]{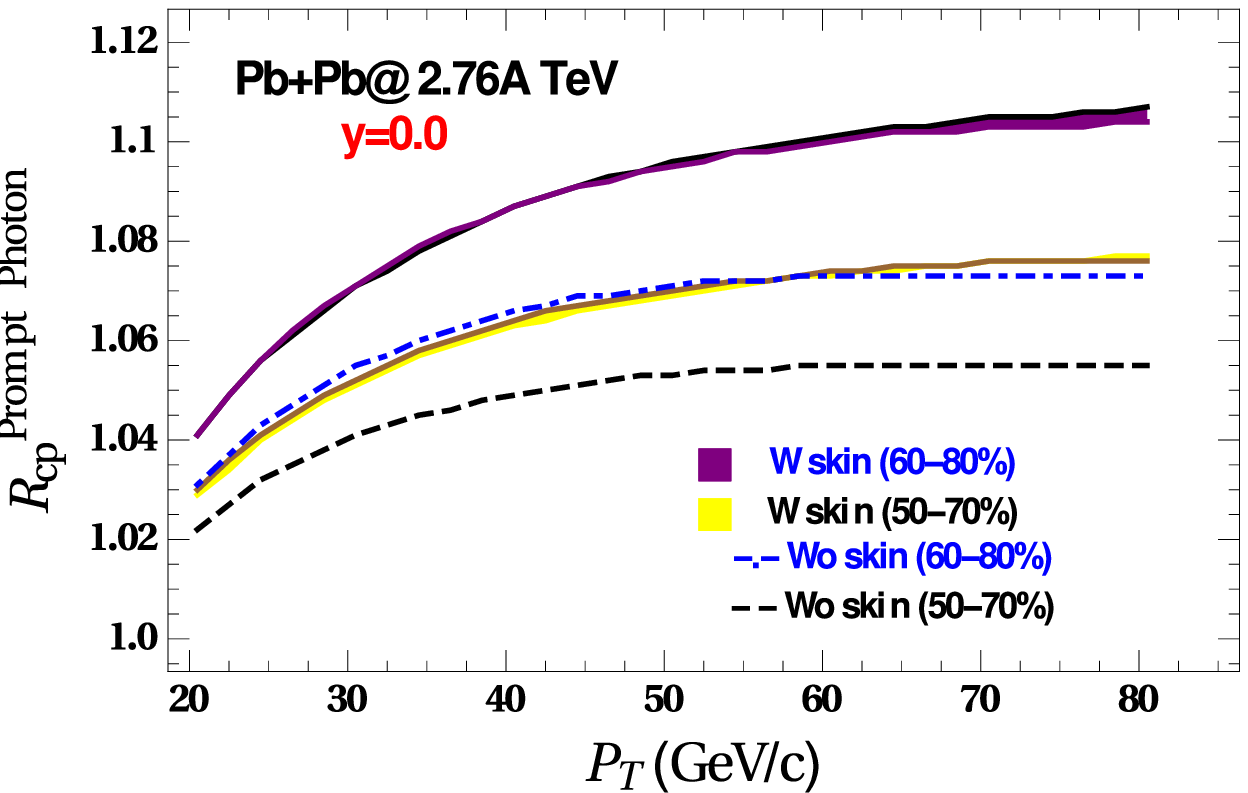}
\includegraphics[width= 7.7cm,clip= true]{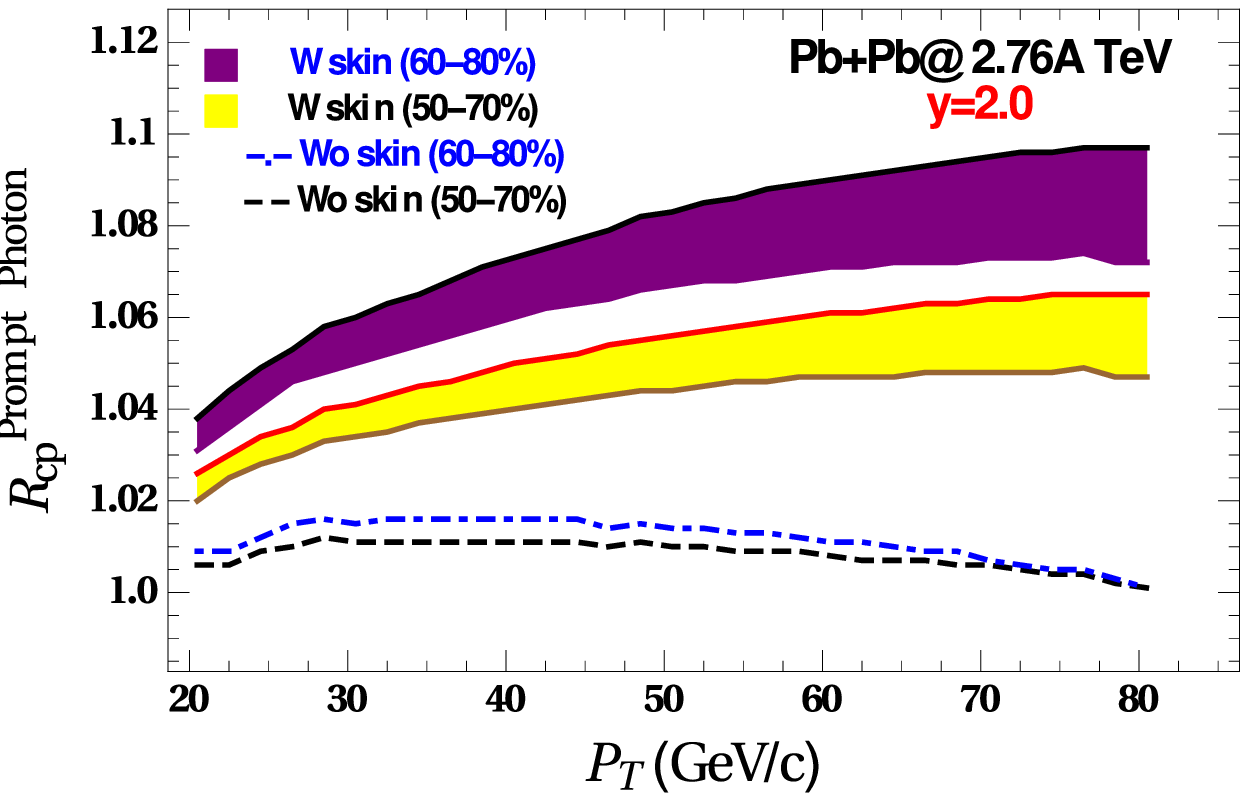}
\end{center}
\caption{(Color online) The \textquoteleft central-to-peripheral ratio\textquoteright ($R_\textrm{cp}$) of prompt photon production is plotted 
against $p_T$ for Pb~+~Pb collisions at $\sqrt{s_{NN}}= 2.76$ TeV at mid-rapidity ($y=$ 0.0) (Left) and 
forward rapidity ($y=$ 2.0) (Right).
The results of \textquoteleft Without\textquoteright neutron skin effect are shown for 50-70\% centrality (black dash) and  
60-80\% centrality (blue dot-dash) of collision. The results of \textquoteleft With\textquoteright neutron skin effect are shown by 
filled bands (light yellow: 50-70\% and dark purple: 60-80\% centrality). The bands depict the 
uncertainty in the estimate, due to error in the parameters of neutron density distribution~\cite{prl14}.}
\label{Rcp-2.76}
\end{figure}
The advantages for this observable are: $R_\textrm{cp}$ has been measured routinely in experiment and the uncertainties in pdf's 
(nuclear and nucleon) relatively cancels out. 
The numerator is evaluated for 0-10\%  centrality and the denominator is calculated for 50-70\% and 60-80\% centrality of Pb~+~Pb collisions. 
In addition, the ratios have been calculated for mid-rapidity ($y=0.0$) and forward rapidity ($y=2.0$) of collisions. \\
To this end we discuss the prompt photon production for Pb~+~Pb collisions at $\sqrt{s_{NN}}=$ 2.76 TeV at three centralities of collision namely,
0-20\%, 20-40\%, 40-80\%. We have used the numerical program INCNLO which evaluates the 
prompt photon production in high energy hadronic collisions up to next-to leading order of strong coupling~\cite{incnlo}.
We have used CTEQ5M1 pdf~\cite{cteq} and BFG-II photon fragmentation function~\cite{BFG} in this work. The factorisation ($Q_f$), 
renormalization ($Q_R$), and fragmentation ($Q_F$) scales are set to a common scale; $Q=p_T$ of the photon. The effective proton 
and neutron numbers for each centrality are 
obtained from Eq.~\ref{zeff}. The results are shown in Fig.~\ref{dirphot} for $p_T>$ 2 GeV/c, along with the direct photon 
data from ALICE Collaboration~\cite{Alice}. It has been found that our results describe the 
data well within statistical uncertainties, specially towards higher $p_T$. We have not included thermal photons from the 
expanding medium (quark-gluon~+~hadron) and jet-medium re-scattering photons in the current study. 
These sources are found greatly important in order to describe the data at low and intermediate $p_T$.
\begin{figure}[t]
\begin{center}
\includegraphics[width= 7.7cm,clip= true]{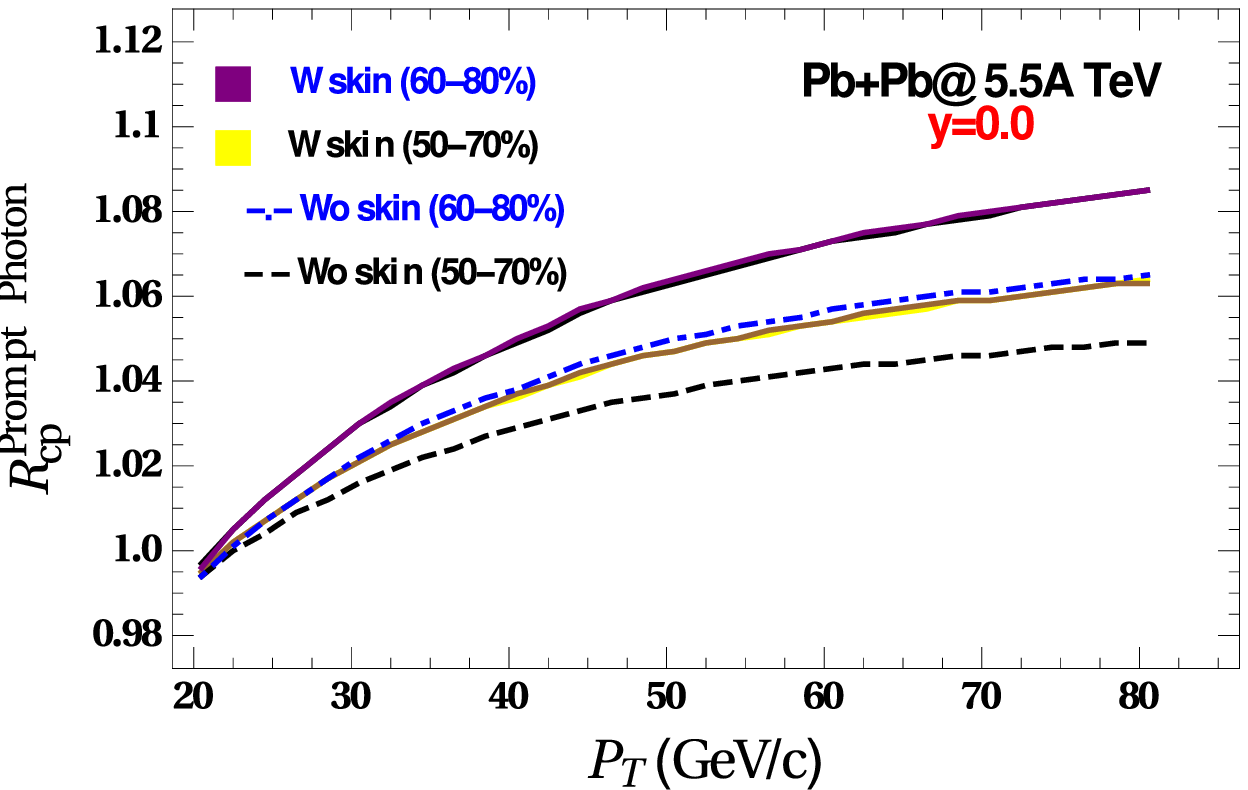}
\includegraphics[width= 7.7cm,clip= true]{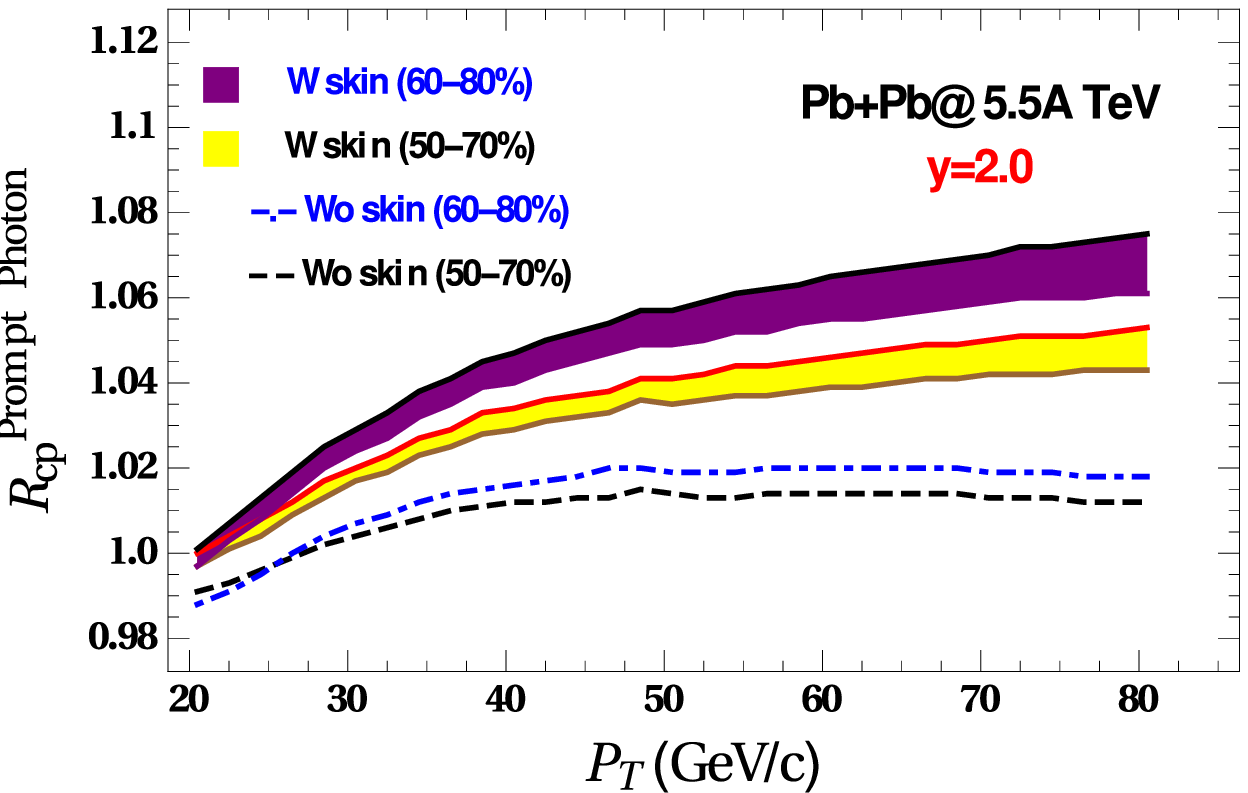}
\end{center}
\caption{(Color online) The \textquoteleft central-to-peripheral ratio\textquoteright ($R_\textrm{cp}$) of prompt photon production 
is plotted against $p_T$ for Pb~+~Pb collisions at $\sqrt{s_{NN}}= 5.5$ TeV at mid-rapidity ($y=$ 0.0) (Left) and 
forward rapidity ($y=$ 2.0) (Right).
The results of \textquoteleft Without\textquoteright neutron skin effect are shown for 50-70\% centrality (black dash) and  
60-80\% centrality (blue dot-dash) of collision. The results of \textquoteleft With\textquoteright neutron skin effect are shown by 
filled bands (light yellow: 50-70\% and dark purple: 60-80\% centrality). The bands depict the 
uncertainty in the estimate, due to error in the parameters of neutron density distribution~\cite{prl14}.}
\label{Rcp-5.5}
\end{figure}

Upon achieving a reliable description of centrality dependent prompt photon production at the LHC energy, we have evaluated 
$R_\textrm{cp}$ of prompt photons starting from $p_T>$ 20 GeV/c. The momentum fraction probed by the prompt
photons of transverse momenta $p_T$ at rapidity $y$: $x\approx (2p_T/\sqrt{s_{NN}}) exp(\pm y)$. Thus we are interested in a regime where 
the photon production probes the valance quark distributions of the initial nucleon pdf. 
The results for Pb~+~Pb collisions at $\sqrt{s_{NN}}=$ 2.76, 5.5 TeV are depicted in Fig.~\ref{Rcp-2.76} and Fig.~\ref{Rcp-5.5} 
respectively, for two rapidities $y=$0.0, 2.0.

Now we shall the discuss the key features of the plots. The momentum fraction probed at mid-rapidity ($y=0.0$) approximately lies 
in the range: 0.01$\lesssim x \lesssim$0.06. The prompt photon production is dominated by gluon distribution of the nucleon wavefunction 
in this regime. The baseline $R_\textrm{cp}$ (i.e., without accounting skin-thickness) is seen to be larger than unity for all values of 
$p_T$. This is owing to the anti-shadowing region of the EKS98 nuclear modification function. The isospin effect turned out to be 
small as the $u$ and $d$ quarks have similar distribution in protons and neutrons. Thus inclusion of neutron skin effect enhances the 
$R_\textrm{cp}$ at the most by$\sim$2-3\% for both collision energies. However at forward rapidity ($y=2.0$), the corresponding momentum fraction 
lies in the range: 0.05$\lesssim x \lesssim$0.5. The behaviour of baseline $R_\textrm{cp}$ is governed by the anti-shadowing and EMC region 
of the nuclear modification function.
Here the prompt photon production is largely sensitive to valance quark distributions of the nucleon wavefunction.  
Hence the neutron skin effect causes a rise in the $R_\textrm{cp}$ by$\sim$7\% for 2.76A TeV 
and by$\sim$5\% for 5.5A TeV center of mass energies on an average. The skin-thickness effect has been found more prominent for 
60-80\% centrality than 50-70\% centrality of collisions for both energies. The band appears in the plots of 
\textquotedblleft W skin\textquotedblright signifies the uncertainty in the photon
production cross-section, arises due to uncertainty in the effective neutron number for a given centrality (see Fig.~\ref{ratio}). 
The increase in width of uncertainty at forward rapidity ascertains the fact that prompt photon production 
is largely isospin dependent in this regime. Also the width is seen to reduce from 2.76A TeV to 5.5A TeV center of mass energy. 
Thus the top LHC energy renders a better opportunity to observe the feature in experiment. 
\section{Summary and Outlook}
In this article, we have discussed the possible implication of neutron skin-thickness of Pb nuclei on inclusive prompt photon production 
in Pb~+~Pb collisions at the LHC energies. Within the scope of optical Glauber model, we found that different nucleon-nucleon scattering 
contributions to the number of participant nucleons are  
separated out in the presence of neutron skin-thickness. The effective proton and neutron numbers are calculated from number of participant nucleons 
using an ansatz. The effective proton to neutron number ratio is found to decrease than the original value in case of large impact parameter 
collisions. We have studied \textquotedblleft central-to-peripheral ratio \textquotedblright ($R_\textrm{cp}$) of prompt photon production 
in order to elucidate the effect. Next-to leading order perturbative QCD has been used to estimate the prompt photon production 
at the LHC energies.  
We have found a discernible enhancement in $R_\textrm{cp}$ while accounting for the neutron skin-thickness. 
This is due to the 
fact that each type of nucleon-nucleon scattering has now contributed to total prompt photon production cross-section.
The enhancement is found to be more prominent at forward rapidity than at mid-rapidity and uncertainty in the estimate reduces towards 
higher center of mass energy. However a comparison with experimental data has not been 
possible due to non-availability of inclusive prompt photon data in this regime. The CMS~\cite{CMS} and ATLAS~\cite{ATLAS} Collaborations have 
excellent measurement capability of large transverse momentum isolated prompt photons at forward rapidity. The isolation criterion necessarily 
cuts down the fragmentation contribution to the inclusive prompt photon cross-section. Thus, in principle, the effect of neutron skin should 
appear in same way in isolated prompt photon production in ultra-peripheral Pb-Pb collisions at the LHC energies. We shall hope to investigate 
the issue in detail in a forthcoming work.\\
\textit{Note added:} We draw attention to the closely related recent article by Helenius {\it et al.}~\cite{nskin-phot1} which had appeared on 
the preprint server while the work was in progress. The authors have also found that the neutron skin effect enhances direct photon production in 
Pb~+~Pb collisions at the LHC energies in a different line of approach.
\section*{Acknowledgement}
SD is thankful to Dinesh Kumar Srivastava for many stimulating discussions regarding the work. The author also thanks Paramita Dutta for her 
suggestions on making of the plots. The author sincerely acknowledges DAE, India for financial support and Grid tier-2 centre, Kolkata for 
providing computing resources during the course of the work. 

 \end{document}